\documentclass[aps,prl,reprint,groupedaddress,showpacs]{revtex4-1}

\usepackage[latin1]{inputenc}
\usepackage{graphicx}
\usepackage{textcomp}
\usepackage{hyperref}
\usepackage{ifthen} 
\usepackage{xifthen} 

\newcommand{\fig}[2][]{%
\ifthenelse{\isempty{#1}}
{Fig.~\ref{#2}}
{Fig.~\ref{#2}(#1)}
}
\newcommand{\Fig}[2][]{%
\ifthenelse{\isempty{#1}}
{Figure~\ref{#2}}
{Figure~\ref{#2}(#1)}
}
\usepackage{color}
\usepackage{soul}

\begin{document}


\title{Signatures of two-level defects in the temperature-dependent damping of nanomechanical silicon nitride resonators}


\author{Thomas Faust}
\author{Johannes Rieger}
\author{Maximilian J. Seitner}
\altaffiliation{Present address: Department of Physics, University of Konstanz, 78457 Konstanz, Germany}
\author{J\"org P. Kotthaus}
\author{Eva M. Weig$^\ast$}
\email{eva.weig@uni-konstanz.de}
\affiliation{Center for NanoScience (CeNS) and Fakult\"at f\"ur Physik, Ludwig-Maximilians-Universit\"at, Geschwister-Scholl-Platz 1,
M\"unchen 80539, Germany}


\begin{abstract}
The damping rates of high quality factor nanomechanical resonators are well beyond intrinsic limits. Here, we explore the underlying microscopic loss mechanisms by investigating the temperature-dependent damping of the fundamental and third harmonic transverse flexural mode of a doubly clamped silicon nitride string. It exhibits characteristic maxima reminiscent of two-level defects typical for amorphous materials. Coupling to those defects relaxes the momentum selection rules, allowing energy transfer from discrete long wavelength resonator modes to the high frequency phonon environment.
\end{abstract}

\maketitle

Silicon nitride (SiN) is a material widely used for resonant micro- and nanomechanical devices because of its superior mechanical properties\,\cite{Sekaric2002,verbridge:124304,bib:Thompson2008a,PhysRevLett.103.207204,Wiederhecker2009,Groblacher2009,Rocheleau2010,Unterreithmeier2010a,PhysRevB.84.165307,PhysRevLett.108.083603,PhysRevB.85.161410,Zhou2013}.
The transverse flexural modes of string resonators fabricated from prestressed SiN thin films exhibit extremely high mechanical quality factors\,\cite{verbridge:124304,Unterreithmeier2010a,PhysRevB.84.165307,PhysRevLett.108.083603}.
They originate from the fact that an increase in tensile stress only slightly increases the mechanical damping rate $\Gamma_m$, whereas it dramatically increases the resonance frequencies $f_m$ and thereby the quality factor $Q=2\pi f_m/\Gamma_m$\,\cite{Unterreithmeier2010a}.
Nonetheless, the observed damping is significantly larger than expected from intrinsic loss mechanisms such as clamping losses caused by the direct radiation of phonons at frequency $f_m$ into the supporting clamping points\,\cite{PhysRevLett.106.047205,Cole2011} or by thermoelastic damping\,\cite{PhysRevB.61.5600}.
In an attempt to shed light on the limiting loss mechanisms, damping was found to be proportional to the local bending within the resonator and governed by both bulk and surface defects\,\cite{Unterreithmeier2010a}.
More recently, the damping has been shown to be dominated by T$_1$-like energy relaxation processes\,\cite{Faust2013}.
Such processes involve a transfer of energy from discrete resonator modes at comparably low frequencies $f_m$ into the high frequency phonon bath that dominates the heat capacity and thermal conductivity.
However, their different dispersion relations inhibit a direct energy transfer via two-particle scattering.
It takes local defects to enable energy transfer into the bath via three particle scattering and to relax momentum conservation.
Such defects are omnipresent in amorphous materials.
For example, a local configurational change of the atomic structure gives rise to a double-well potential separated by an energy barrier, which at low temperatures can be modeled as a two-level system (TLS).
These TLS are known to lead to characteristic maxima in the temperature dependence of the sound absorption in the temperature range of 10 to 100\,K\,\cite{PhysRevB.45.2750,RevModPhys.74.991,PhysRevB.72.214205,PhysRevLett.96.055902}.
The signature of TLS has also been observed in the damping characteristics of a micromechanical silica resonator\,\cite{PhysRevA.80.021803} and in a backaction-evading measurement on a SiN membrane performed at mK temperatures\,\cite{suh:052604}. 

To clarify whether the microscopic nature of the damping in high Q SiN nano resonators is dominated by local defect scattering induced by such two-level systems, we study the temperature-dependent damping of nanoscale string resonators fabricated from prestressed SiN films.
Simultaneous measurements of the fundamental and third harmonic flexural mode of such a resonator allow to not only test the characteristic temperature- but also the frequency-dependence of the established TLS model.
Our findings demonstrate that the two-level defect states thus found in silicon nitride are rather similar to the ones found in silica\,\cite{PhysRevB.72.214205} and amorphous silicon\,\cite{PhysRevLett.96.055902}.
In contrast to high-purity silica, our SiN resonators feature a second maximum in the temperature-dependent damping, which might be attributed to hydrogen contamination during thin film deposition.
An only weakly temperature-dependent damping background most likely originates from damping via surface defects.
Our findings will enable further increase of quality factors of nanomechanical resonators towards intrinsic limits set by the energy transfer to the environment\,\cite{PhysRevLett.106.047205,Cole2011,PhysRevB.61.5600}.
For the fundamental string resonator mode at a frequency of about $6.8$\,MHz studied below, the dominant limit, set by clamping losses, is estimated to be about 3 million\,\cite{Wilson-Rae2013} at room temperature, an order of magnitude above the observed Q of 0.3 million.

The nanomechanical resonator exemplarily shown in this work is a 100\,nm thick, 250\,nm wide and 55\,\textmu m long doubly clamped string fabricated from a prestressed silicon nitride film deposited on a fused silica substrate\,\cite{Faust2012}.
It is flanked by two lower-lying gold electrodes which are used for gradient-field induced dielectric actuation\,\cite{Unterreithmeier2009} and simultaneously\,\cite{rieger:103110} allow for sensitive electrical detection of its motion via the coupling to a microwave cavity\,\cite{Faust2012} with frequency $f_c\gg f_m$ $(m=1,3)$.
The mechanical resonator is investigated at pressures below $5\cdot10^{-5}$\,mbar inside a pulse tube cooler\,\cite{Faust2013} to avoid gas damping.
Control of the pulse tube operation in combination with a powerful Ohmic heater allows to stabilize the temperature of the sample (measured inside the brass sample holder a few millimeters below the fused silica chip) anywhere between 7 and 350\,K with a precision of at least 0.1\,K.

\begin{figure}
\includegraphics{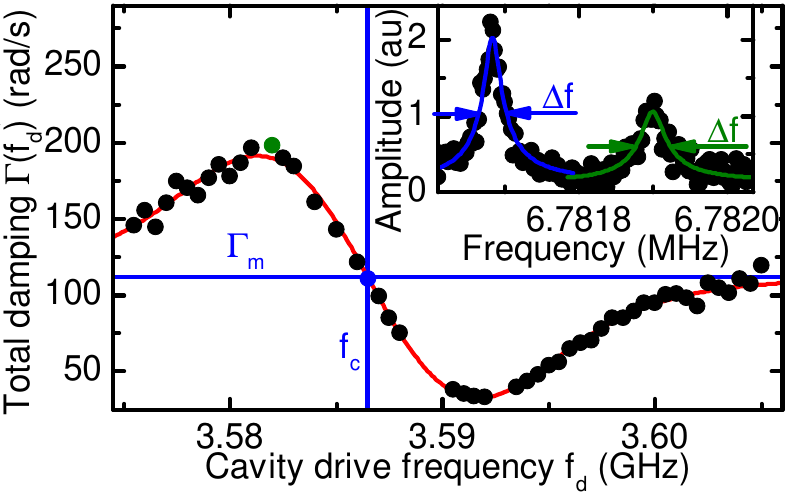}
\caption{\label{optomech}(color online). 
Total mechanical damping of the fundamental resonator mode versus microwave cavity drive frequency (black dots) at 10.5\,K. The total damping $\Gamma=2\pi \Delta f$ consists of the intrinsic mechanical damping $\Gamma_m$ and an optomechanical damping contribution $\Gamma_{\rm opt}$. By fitting the total damping versus cavity drive frequency curve (red line), the intrinsic mechanical damping $\Gamma_{\rm m}$ $(m=1)$ and the cavity resonance frequency $f_{\rm c}$ can be extracted for the respective temperature. Two exemplary spectra with Lorentzian fits, corresponding to maximal (right, green) and zero (left, blue) optomechanical damping (see colored dots in the main figure), are shown in the inset.}
\end{figure}

The microwave detection scheme exerts radiation pressure induced optomechanical backaction forces on the mechanical resonator\,\cite{1367-2630-10-9-095002,Faust2012}.
This causes the measured damping $\Gamma=2\pi \Delta f$, extracted from the linewidth $\Delta f$ of the mechanical resonance, to characteristically depend on the detuning of the microwave drive frequency $f_d$ from the cavity resonance frequency $f_c$\,\cite{Faust2012}.
We extract the intrinsic damping $\Gamma_m$ $(m=1,3)$ at $f_d=f_c$ where backaction effects are negligible, as shown in \fig{optomech}.
To account for the temperature dependence of the cavity resonance, mechanical resonance curves are recorded for a multitude of cavity drive frequencies $f_d$ at every temperature.
This measurement is conducted in parallel for the fundamental out-of-plane mode with a resonance frequency $f_1$=6.8\,MHz and the third harmonic mode at $f_3$=20.2\,MHz using two network analyzers multiplexed via powersplitters.
Exemplarily, \fig{optomech} shows the response of the fundamental mode at 10.5\,K.
A fit of the optomechanically influenced\,\cite{1367-2630-10-9-095002,Faust2012} total damping $\Gamma(f_d)=\Gamma_m+\Gamma_{\rm opt}(f_d)$ (corresponding to the linewidth $\Delta f$, see inset) yields the net intrinsic damping $\Gamma_1$ of the fundamental mode and also the precise cavity resonance frequency $f_c$ at this temperature.
The intrinsic damping of the third harmonic mode $\Gamma_3$ is obtained by taking the average over the 20 closest datapoints to the cavity resonance frequency $f_c$.
A direct fit is not possible as the optomechanical damping of this mode is too weak, however, it is point symmetric around $f_c$ (cf. \fig{optomech}) such that averaging yields the intrinsic $\Gamma_3$.

The temperature of the cryostat is now varied between 7.5 and 349.5\,K in steps of 1\,K, and the above procedure is repeated for every temperature.
\Fig[a]{data} shows the obtained temperature dependence of the intrinsic damping $\Gamma_1(T)$ of the fundamental mode with error bars indicating the errors extracted from the optomechanical fits.
The damping of the third harmonic $\Gamma_3(T)$ is plotted versus temperature in \fig[b]{data}, here the error bars correspond to the standard deviation of the 20 data points closest to $f_c$.
Both spectra similarly exhibit two distinct maxima, one at approximately 50\,K and the other one near 200\,K.

\begin{figure}
\includegraphics{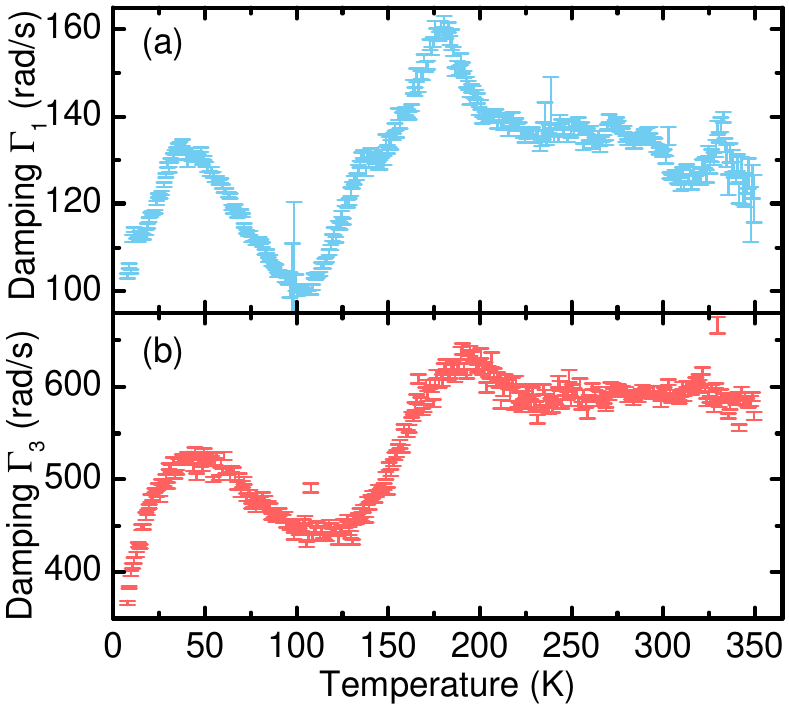}
\caption{\label{data}(color online).
The measured mode damping $\Gamma_1$ of the fundamental mode and $\Gamma_3$ of the third harmonic mode is shown versus temperature in panels (a) and (b), respectively. The error bars in (a) represent the errors of the optomechanical fits like the one shown in \fig{optomech} while the ones in (b) correspond to the standard deviation of the averaged measurements.}
\end{figure}

\begin{figure*}
\includegraphics{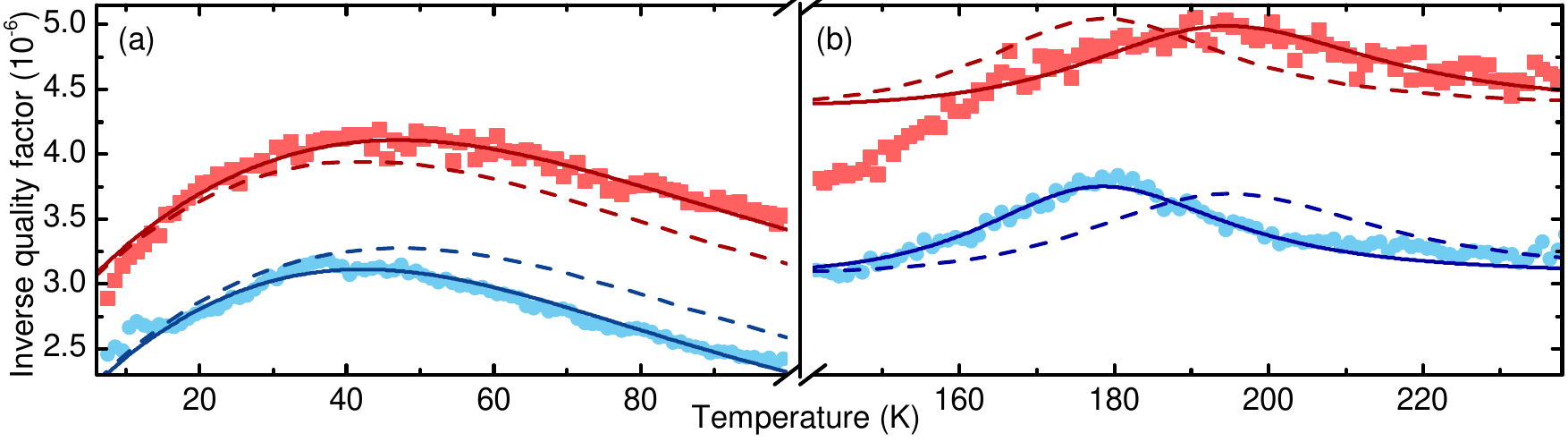}
\caption{\label{fits}(color online). Fitted TLS (a) and Arrhenius (b) peaks of the fundamental (lower, blue) and third harmonic (upper, red) mode. Both classes of peaks were fitted using a single set of parameters for both modes except a mode-dependent damping background. The actual fits are shown as solid lines. The dashed lines illustrate the frequency dependence of the fits as they use the same set of parameters, but interchanged frequencies of the two modes.}
\end{figure*}

The lower-temperature peak is a clear signature of the two-level system defects in the amorphous SiN\,\cite{PhysRevB.72.214205}.
These configurational changes in a glassy material can be modeled by a particle in a double-well potential with barrier height $V$ and asymmetry $\Delta$ between the two wells which tries to overcome the barrier with an attempt frequency $\tau_0^{-1}$.
As there are a multitude of different defects, both $V$ and $\Delta$ are distributed over a certain range, characterized by two cutoff parameters $V_0$ and $\Delta_C$\,\cite{PhysRevB.72.214205}.
At the elevated temperatures studied in this experiment, thermally activated processes dominate over resonant tunneling, leading to a pronounced maximum in the damping rate when the hopping rate, i.\,e. the inverse of $\tau=\tau_0e^{V/T}$ ($V$ and $\Delta$ are always given in units of temperature) is equal to the mechanical oscillation frequency $2\pi f_m$ of the respective mode.
Assuming a mode-dependent damping background, probably caused by surface defects, the low-temperature peaks of both modes can be fitted with a single set of parameters employing equation (9a) of Ref.\,\cite{PhysRevB.72.214205}.
To this end, the measured damping constants are converted into an inverse quality factor $Q^{-1}=\Gamma_m/(2\pi f_m)$, which is the commonly used quantity in the field of TLS damping.
Note, however, that inverse quality factors can only be used to compare the internal friction of resonators with equal stress, as tensile stress increases $f_m$ with little change in $\Gamma_m$\,\cite{Unterreithmeier2010a}.
The data points of both modes and the respective fits\,\cite{PhysRevB.72.214205} (solid lines) are shown in \fig[a]{fits}.
The dashed lines, which are obtained by plotting the resulting fit functions with the two mechanical frequencies exchanged, illustrate the clear frequency dependence of the fit.
The parameters extracted from the fit are $V_0=460\pm4$\,K, $\Lambda_C=110\pm2$\,K and $\log_{10}\tau_0/\rm{s}=-11.24\pm0.02$ using a background of $1.78\pm0.02\cdot 10^{-6}$ and $2.62\pm0.02\cdot 10^{-6}$ for the fundamental and third harmonic mode, respectively.
Comparing these values with the ones reported for silica\,\cite{PhysRevB.72.214205} for a lack of reference values on SiN, a maximum deviation of 30\% is observed, indicating that the nature of the defect states is rather similar in both materials.

The dissipation maximum found at temperatures near 200\,K can be explained by a so-called Arrhenius peak\,\cite{PhysRev.123.1204,Nowick1972,Kappesser1996,hutchinson:972,Martin2010}, indicative of a thermally activated relaxation process over a well-defined barrier height $V$ (in constrast to the broad distribution governing the behaviour of the typical two-level defects in glassy material).
The most likely candidate for these well-defined states might be the hydrogen defects present even in very clean LPCVD silicon nitride films\,\cite{chow:5630}.
The dissipation behaviour of such a defect can be modeled assuming a delta-function distribution located at $V=V_A$ for the barrier height and by neglecting the comparatively small energy difference between the two potential wells.
Both peaks can be fitted using the same set of parameters but two somewhat different background damping constants for the two modes.
The measured inverse quality factors along with the solid fit lines are shown in \fig[b]{fits}.
As in panel (a), the dashed lines demonstrate the frequency dependence of the fit by illustrating the fitting functions with interchanged mechanical frequencies.
A barrier height of $V_a=2354\pm10$\,K and an inverse attempt frequency $\log_{10}\tau_a/\rm{s}=-13.32\pm0.02$ are extracted for the Arrhenius peaks, using a dissipation background of $3.08\pm0.02\cdot10^{-6}$ and $4.37\pm0.02\cdot10^{-6}$ for fundamental and third harmonic mode, respectively.

Comparison of the dissipation backgrounds found at low and high temperatures reveals a slow increase of the overall damping with increasing temperature for both modes.
This is consistent with earlier studies\,\cite{hutchinson:972,Shimada1984}, and may be connected to the temperature dependent damping of surface defects (see Supplement of \cite{Unterreithmeier2010a}), which are the most likely candidates to cause the remaining dissipation not accounted for in our analysis.

\begin{figure}
\includegraphics{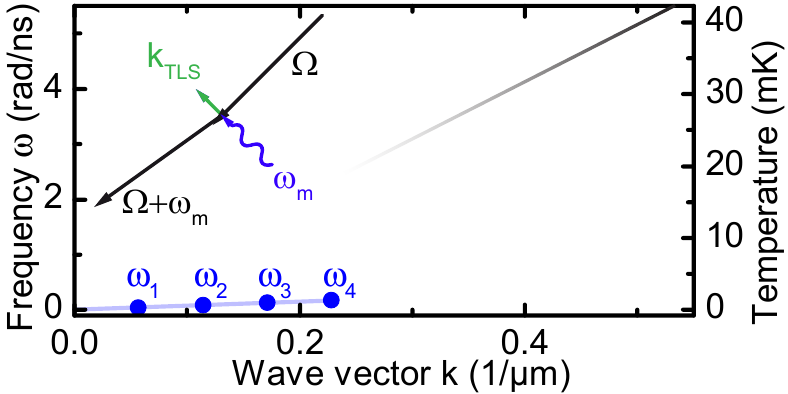}
\caption{\label{theo}(color online). Dispersion relation of the discrete modes of the nanomechanical resonator (blue or gray) alongside the longitudinal bulk phonon mode visualizing the situation inside the string (black). The mechanical impedance mismatch between resonator and clamping points makes it increasingly difficult for long-wavelength bulk phonons to enter or leave the mechanical resonator, which is indicated by the fading black line. The inset visualizes a scattering process involving a two-level defect system which allows a thermally excited high-energy phonon $\Omega$ to absorb the energy of one resonator phonon $\omega_m=2\pi f_m$ and carry it out of the string.}
\end{figure}

The experimental data convincingly demonstrates that material defects play a significant role in the damping of prestressed SiN nanoresonators.
In the following, we will establish a corresponding microscopic picture which consistently models the experimental evidence that damping is (a) caused by energy relaxation processes\,\cite{Faust2013}, (b) involves TLS and (c) is not limited by the well-known intrinsic damping mechanisms\,\cite{PhysRevLett.106.047205,Cole2011,PhysRevB.61.5600}.
For the sake of simplicity we will employ a quasiparticle approach based on phonon scattering to illustrate how mechanical vibration energy is transported out of the string during each oscillation period:
The dispersion relation of the longitudinal bulk phonon mode\,\cite{holmes:2250} along with the discrete flexural resonator modes are shown in \fig{theo}.
Bulk modes with a small wave vector $k$ can not enter or leave the mechanical resonator as they are reflected at the huge mechanical impedance mismatch at the clamping ponts\,\cite{Rieger2013}, making them play a negligible role in the energy transport out of the string.
The only way for the two kinds of modes to interact is via localized defect states inside the resonator, which mediate the interaction between the freely propagating high-energy bulk phonons and the low-energy resonator phonons.
They also provide the excess momentum to enable thermally excited bulk phonons to scatter off a resonator phonon.
This process is illustrated by the inset sketch in \fig{theo}.
As the high-energy bulk phonon modes are highly populated at temperatures of a few kelvin (see the temperature scale in \fig{theo}), all resonator phonons which interact with a TLS are not re-emitted into the mode but rather scatter with a bulk phonon.
Hence, the proposed process is consistent with the observation that there is no measurable phase relaxation of the mechanical mode\,\cite{Faust2013}.
Furthermore, high-energy bulk phonons with velocity $v_{\rm ph}$ and mean free path $l_{\rm ph}$ can efficiently remove the excess energy $\omega_m$ from the resonator as long as their thermal relaxation rate $\Gamma_{\rm th}\gg\Gamma_m$.
We estimate $\Gamma_{\rm th}=G_{\rm th}/C\simeq v_{\rm ph}l_{\rm ph}/L^2$, governed by the thermal conductance $G_{\rm th}$ and the heat capacitance $C$ of the resonator of length $L$, to be at least an order of magnitude faster than $\Gamma_m$ with $l_{\rm ph}$ of a few nm in glassy materials\,\cite{holmes:2250,PhysRevLett.96.055902}.
Thus, three particle scattering with a TLS can indeed provide the required energy relaxation paths to damp the resonator vibration.

In conclusion, our results demonstrate that the internal dissipation of a nanomechanical SiN resonator is governed by the well-established microscopic phenomena known from low-temperature glass physics.
The temperature-dependent damping reveals that two-level system defects, i.\,e. configurational changes of the atomic structure in an amorphous material, are also found in silicon nitride.
Additionaly, a second, Arrhenius-type peak in the damping observed at higher temperatures indicates another kind of impurity in the material, probably hydrogen atoms incorporated during SiN thin film deposition.

The presented measurements offer a new way to analyze the mechanical quality of SiN films and may even be useful to quantify the hydrogen content.
Furthermore, the observation that these dissipation mechanisms are related to the amorphous structure of the material clearly sets expectations for even higher quality factors in single crystalline resonators if subjected to similar tensile stress.
Assuming a negligible influence of surface defects caused by fabrication or subsequent oxidation, their quality factor should only be limited by fundamental effects such as clamping losses and thermoelastic damping, which leaves room for improvement by at least an order of magnitude.

\begin{acknowledgments}
Financial support by the Deutsche Forschungsgemeinschaft via Project No. Ko 416/18, the German Excellence Initiative via the Nanosystems Initiative Munich (NIM) and LMUexcellent, as well as the European Commission under the FET-Open project QNEMS (233992) is gratefully acknowledged. We thank S. Ludwig for his insightful comments on the Arrhenius peak.
\end{acknowledgments}

\bibliography{../Dissertation_Faust/paper}

\begin{thebibliography}{36}%
\makeatletter
\providecommand \@ifxundefined [1]{%
 \@ifx{#1\undefined}
}%
\providecommand \@ifnum [1]{%
 \ifnum #1\expandafter \@firstoftwo
 \else \expandafter \@secondoftwo
 \fi
}%
\providecommand \@ifx [1]{%
 \ifx #1\expandafter \@firstoftwo
 \else \expandafter \@secondoftwo
 \fi
}%
\providecommand \natexlab [1]{#1}%
\providecommand \enquote  [1]{``#1''}%
\providecommand \bibnamefont  [1]{#1}%
\providecommand \bibfnamefont [1]{#1}%
\providecommand \citenamefont [1]{#1}%
\providecommand \href@noop [0]{\@secondoftwo}%
\providecommand \href [0]{\begingroup \@sanitize@url \@href}%
\providecommand \@href[1]{\@@startlink{#1}\@@href}%
\providecommand \@@href[1]{\endgroup#1\@@endlink}%
\providecommand \@sanitize@url [0]{\catcode `\\12\catcode `\$12\catcode
  `\&12\catcode `\#12\catcode `\^12\catcode `\_12\catcode `\%12\relax}%
\providecommand \@@startlink[1]{}%
\providecommand \@@endlink[0]{}%
\providecommand \url  [0]{\begingroup\@sanitize@url \@url }%
\providecommand \@url [1]{\endgroup\@href {#1}{\urlprefix }}%
\providecommand \urlprefix  [0]{URL }%
\providecommand \Eprint [0]{\href }%
\providecommand \doibase [0]{http://dx.doi.org/}%
\providecommand \selectlanguage [0]{\@gobble}%
\providecommand \bibinfo  [0]{\@secondoftwo}%
\providecommand \bibfield  [0]{\@secondoftwo}%
\providecommand \translation [1]{[#1]}%
\providecommand \BibitemOpen [0]{}%
\providecommand \bibitemStop [0]{}%
\providecommand \bibitemNoStop [0]{.\EOS\space}%
\providecommand \EOS [0]{\spacefactor3000\relax}%
\providecommand \BibitemShut  [1]{\csname bibitem#1\endcsname}%
\let\auto@bib@innerbib\@empty
\bibitem [{\citenamefont {Sekaric}\ \emph {et~al.}(2002)\citenamefont
  {Sekaric}, \citenamefont {Carr}, \citenamefont {Evoy}, \citenamefont
  {Parpia},\ and\ \citenamefont {Craighead}}]{Sekaric2002}%
  \BibitemOpen
  \bibfield  {author} {\bibinfo {author} {\bibfnamefont {L.}~\bibnamefont
  {Sekaric}}, \bibinfo {author} {\bibfnamefont {D.~W.}\ \bibnamefont {Carr}},
  \bibinfo {author} {\bibfnamefont {S.}~\bibnamefont {Evoy}}, \bibinfo {author}
  {\bibfnamefont {J.~M.}\ \bibnamefont {Parpia}}, \ and\ \bibinfo {author}
  {\bibfnamefont {H.~G.}\ \bibnamefont {Craighead}},\ }\href {\doibase
  http://dx.doi.org/10.1016/S0924-4247(02)00149-8} {\bibfield  {journal}
  {\bibinfo  {journal} {Sensor. Actuat. A-Phys.}\ }\textbf {\bibinfo {volume}
  {101}},\ \bibinfo {pages} {215 } (\bibinfo {year} {2002})}\BibitemShut
  {NoStop}%
\bibitem [{\citenamefont {Verbridge}\ \emph {et~al.}(2006)\citenamefont
  {Verbridge}, \citenamefont {Parpia}, \citenamefont {Reichenbach},
  \citenamefont {Bellan},\ and\ \citenamefont {Craighead}}]{verbridge:124304}%
  \BibitemOpen
  \bibfield  {author} {\bibinfo {author} {\bibfnamefont {S.~S.}\ \bibnamefont
  {Verbridge}}, \bibinfo {author} {\bibfnamefont {J.~M.}\ \bibnamefont
  {Parpia}}, \bibinfo {author} {\bibfnamefont {R.~B.}\ \bibnamefont
  {Reichenbach}}, \bibinfo {author} {\bibfnamefont {L.~M.}\ \bibnamefont
  {Bellan}}, \ and\ \bibinfo {author} {\bibfnamefont {H.~G.}\ \bibnamefont
  {Craighead}},\ }\href {\doibase 10.1063/1.2204829} {\bibfield  {journal}
  {\bibinfo  {journal} {J. Appl. Phys.}\ }\textbf {\bibinfo {volume} {99}},\
  \bibinfo {eid} {124304} (\bibinfo {year} {2006})}\BibitemShut {NoStop}%
\bibitem [{\citenamefont {Thompson}\ \emph {et~al.}(2008)\citenamefont
  {Thompson}, \citenamefont {Zwickl}, \citenamefont {Jayich}, \citenamefont
  {Marquardt}, \citenamefont {Girvin},\ and\ \citenamefont
  {Harris}}]{bib:Thompson2008a}%
  \BibitemOpen
  \bibfield  {author} {\bibinfo {author} {\bibfnamefont {J.~D.}\ \bibnamefont
  {Thompson}}, \bibinfo {author} {\bibfnamefont {B.~M.}\ \bibnamefont
  {Zwickl}}, \bibinfo {author} {\bibfnamefont {A.~M.}\ \bibnamefont {Jayich}},
  \bibinfo {author} {\bibfnamefont {F.}~\bibnamefont {Marquardt}}, \bibinfo
  {author} {\bibfnamefont {S.~M.}\ \bibnamefont {Girvin}}, \ and\ \bibinfo
  {author} {\bibfnamefont {J.~G.~E.}\ \bibnamefont {Harris}},\ }\href {\doibase
  10.1038/nature06715} {\bibfield  {journal} {\bibinfo  {journal} {Nature}\
  }\textbf {\bibinfo {volume} {452}},\ \bibinfo {pages} {72} (\bibinfo {year}
  {2008})}\BibitemShut {NoStop}%
\bibitem [{\citenamefont {Wilson}\ \emph {et~al.}(2009)\citenamefont {Wilson},
  \citenamefont {Regal}, \citenamefont {Papp},\ and\ \citenamefont
  {Kimble}}]{PhysRevLett.103.207204}%
  \BibitemOpen
  \bibfield  {author} {\bibinfo {author} {\bibfnamefont {D.~J.}\ \bibnamefont
  {Wilson}}, \bibinfo {author} {\bibfnamefont {C.~A.}\ \bibnamefont {Regal}},
  \bibinfo {author} {\bibfnamefont {S.~B.}\ \bibnamefont {Papp}}, \ and\
  \bibinfo {author} {\bibfnamefont {H.~J.}\ \bibnamefont {Kimble}},\ }\href
  {\doibase 10.1103/PhysRevLett.103.207204} {\bibfield  {journal} {\bibinfo
  {journal} {Phys. Rev. Lett.}\ }\textbf {\bibinfo {volume} {103}},\ \bibinfo
  {pages} {207204} (\bibinfo {year} {2009})}\BibitemShut {NoStop}%
\bibitem [{\citenamefont {Wiederhecker}\ \emph {et~al.}(2009)\citenamefont
  {Wiederhecker}, \citenamefont {Chen}, \citenamefont {Gondarenko},\ and\
  \citenamefont {Lipson}}]{Wiederhecker2009}%
  \BibitemOpen
  \bibfield  {author} {\bibinfo {author} {\bibfnamefont {G.~S.}\ \bibnamefont
  {Wiederhecker}}, \bibinfo {author} {\bibfnamefont {L.}~\bibnamefont {Chen}},
  \bibinfo {author} {\bibfnamefont {A.}~\bibnamefont {Gondarenko}}, \ and\
  \bibinfo {author} {\bibfnamefont {M.}~\bibnamefont {Lipson}},\ }\href
  {\doibase 10.1038/nature08584} {\bibfield  {journal} {\bibinfo  {journal}
  {Nature}\ }\textbf {\bibinfo {volume} {462}},\ \bibinfo {pages} {633}
  (\bibinfo {year} {2009})}\BibitemShut {NoStop}%
\bibitem [{\citenamefont {{Gröblacher}}\ \emph {et~al.}(2009)\citenamefont
  {{Gröblacher}}, \citenamefont {Hertzberg}, \citenamefont {Vanner},
  \citenamefont {Cole}, \citenamefont {Gigan}, \citenamefont {Schwab},\ and\
  \citenamefont {Aspelmeyer}}]{Groblacher2009}%
  \BibitemOpen
  \bibfield  {author} {\bibinfo {author} {\bibfnamefont {S.}~\bibnamefont
  {{Gröblacher}}}, \bibinfo {author} {\bibfnamefont {J.~B.}\ \bibnamefont
  {Hertzberg}}, \bibinfo {author} {\bibfnamefont {M.~R.}\ \bibnamefont
  {Vanner}}, \bibinfo {author} {\bibfnamefont {G.~D.}\ \bibnamefont {Cole}},
  \bibinfo {author} {\bibfnamefont {S.}~\bibnamefont {Gigan}}, \bibinfo
  {author} {\bibfnamefont {K.~C.}\ \bibnamefont {Schwab}}, \ and\ \bibinfo
  {author} {\bibfnamefont {M.}~\bibnamefont {Aspelmeyer}},\ }\href
  {http://dx.doi.org/10.1038/nphys1301} {\bibfield  {journal} {\bibinfo
  {journal} {Nat Phys}\ }\textbf {\bibinfo {volume} {5}},\ \bibinfo {pages}
  {485} (\bibinfo {year} {2009})}\BibitemShut {NoStop}%
\bibitem [{\citenamefont {Rocheleau}\ \emph {et~al.}(2010)\citenamefont
  {Rocheleau}, \citenamefont {Ndukum}, \citenamefont {Macklin}, \citenamefont
  {Hertzberg}, \citenamefont {Clerk},\ and\ \citenamefont
  {Schwab}}]{Rocheleau2010}%
  \BibitemOpen
  \bibfield  {author} {\bibinfo {author} {\bibfnamefont {T.}~\bibnamefont
  {Rocheleau}}, \bibinfo {author} {\bibfnamefont {T.}~\bibnamefont {Ndukum}},
  \bibinfo {author} {\bibfnamefont {C.}~\bibnamefont {Macklin}}, \bibinfo
  {author} {\bibfnamefont {J.~B.}\ \bibnamefont {Hertzberg}}, \bibinfo {author}
  {\bibfnamefont {A.~A.}\ \bibnamefont {Clerk}}, \ and\ \bibinfo {author}
  {\bibfnamefont {K.~C.}\ \bibnamefont {Schwab}},\ }\href
  {http://dx.doi.org/10.1038/nature08681} {\bibfield  {journal} {\bibinfo
  {journal} {Nature}\ }\textbf {\bibinfo {volume} {463}},\ \bibinfo {pages}
  {72} (\bibinfo {year} {2010})}\BibitemShut {NoStop}%
\bibitem [{\citenamefont {{Unterreithmeier}}\ \emph {et~al.}(2010)\citenamefont
  {{Unterreithmeier}}, \citenamefont {{Faust}},\ and\ \citenamefont
  {{Kotthaus}}}]{Unterreithmeier2010a}%
  \BibitemOpen
  \bibfield  {author} {\bibinfo {author} {\bibfnamefont {Q.~P.}\ \bibnamefont
  {{Unterreithmeier}}}, \bibinfo {author} {\bibfnamefont {T.}~\bibnamefont
  {{Faust}}}, \ and\ \bibinfo {author} {\bibfnamefont {J.~P.}\ \bibnamefont
  {{Kotthaus}}},\ }\href {\doibase 10.1103/PhysRevLett.105.027205} {\bibfield
  {journal} {\bibinfo  {journal} {Phys. Rev. Lett.}\ }\textbf {\bibinfo
  {volume} {105}},\ \bibinfo {eid} {027205} (\bibinfo {year}
  {2010})}\BibitemShut {NoStop}%
\bibitem [{\citenamefont {Schmid}\ \emph {et~al.}(2011)\citenamefont {Schmid},
  \citenamefont {Jensen}, \citenamefont {Nielsen},\ and\ \citenamefont
  {Boisen}}]{PhysRevB.84.165307}%
  \BibitemOpen
  \bibfield  {author} {\bibinfo {author} {\bibfnamefont {S.}~\bibnamefont
  {Schmid}}, \bibinfo {author} {\bibfnamefont {K.~D.}\ \bibnamefont {Jensen}},
  \bibinfo {author} {\bibfnamefont {K.~H.}\ \bibnamefont {Nielsen}}, \ and\
  \bibinfo {author} {\bibfnamefont {A.}~\bibnamefont {Boisen}},\ }\href
  {\doibase 10.1103/PhysRevB.84.165307} {\bibfield  {journal} {\bibinfo
  {journal} {Phys. Rev. B}\ }\textbf {\bibinfo {volume} {84}},\ \bibinfo
  {pages} {165307} (\bibinfo {year} {2011})}\BibitemShut {NoStop}%
\bibitem [{\citenamefont {Yu}\ \emph {et~al.}(2012)\citenamefont {Yu},
  \citenamefont {Purdy},\ and\ \citenamefont {Regal}}]{PhysRevLett.108.083603}%
  \BibitemOpen
  \bibfield  {author} {\bibinfo {author} {\bibfnamefont {P.-L.}\ \bibnamefont
  {Yu}}, \bibinfo {author} {\bibfnamefont {T.~P.}\ \bibnamefont {Purdy}}, \
  and\ \bibinfo {author} {\bibfnamefont {C.~A.}\ \bibnamefont {Regal}},\ }\href
  {\doibase 10.1103/PhysRevLett.108.083603} {\bibfield  {journal} {\bibinfo
  {journal} {Phys. Rev. Lett.}\ }\textbf {\bibinfo {volume} {108}},\ \bibinfo
  {pages} {083603} (\bibinfo {year} {2012})}\BibitemShut {NoStop}%
\bibitem [{\citenamefont {Fong}\ \emph {et~al.}(2012)\citenamefont {Fong},
  \citenamefont {Pernice},\ and\ \citenamefont {Tang}}]{PhysRevB.85.161410}%
  \BibitemOpen
  \bibfield  {author} {\bibinfo {author} {\bibfnamefont {K.~Y.}\ \bibnamefont
  {Fong}}, \bibinfo {author} {\bibfnamefont {W.~H.~P.}\ \bibnamefont
  {Pernice}}, \ and\ \bibinfo {author} {\bibfnamefont {H.~X.}\ \bibnamefont
  {Tang}},\ }\href {\doibase 10.1103/PhysRevB.85.161410} {\bibfield  {journal}
  {\bibinfo  {journal} {Phys. Rev. B}\ }\textbf {\bibinfo {volume} {85}},\
  \bibinfo {pages} {161410} (\bibinfo {year} {2012})}\BibitemShut {NoStop}%
\bibitem [{\citenamefont {Zhou}\ \emph {et~al.}(2013)\citenamefont {Zhou},
  \citenamefont {Hocke}, \citenamefont {Schliesser}, \citenamefont {Marx},
  \citenamefont {Huebl}, \citenamefont {Gross},\ and\ \citenamefont
  {Kippenberg}}]{Zhou2013}%
  \BibitemOpen
  \bibfield  {author} {\bibinfo {author} {\bibfnamefont {X.}~\bibnamefont
  {Zhou}}, \bibinfo {author} {\bibfnamefont {F.}~\bibnamefont {Hocke}},
  \bibinfo {author} {\bibfnamefont {A.}~\bibnamefont {Schliesser}}, \bibinfo
  {author} {\bibfnamefont {A.}~\bibnamefont {Marx}}, \bibinfo {author}
  {\bibfnamefont {H.}~\bibnamefont {Huebl}}, \bibinfo {author} {\bibfnamefont
  {R.}~\bibnamefont {Gross}}, \ and\ \bibinfo {author} {\bibfnamefont {T.~J.}\
  \bibnamefont {Kippenberg}},\ }\href {http://dx.doi.org/10.1038/nphys2527}
  {\bibfield  {journal} {\bibinfo  {journal} {Nat Phys}\ }\textbf {\bibinfo
  {volume} {9}},\ \bibinfo {pages} {179} (\bibinfo {year} {2013})}\BibitemShut
  {NoStop}%
\bibitem [{\citenamefont {Wilson-Rae}\ \emph {et~al.}(2011)\citenamefont
  {Wilson-Rae}, \citenamefont {Barton}, \citenamefont {Verbridge},
  \citenamefont {Southworth}, \citenamefont {Ilic}, \citenamefont {Craighead},\
  and\ \citenamefont {Parpia}}]{PhysRevLett.106.047205}%
  \BibitemOpen
  \bibfield  {author} {\bibinfo {author} {\bibfnamefont {I.}~\bibnamefont
  {Wilson-Rae}}, \bibinfo {author} {\bibfnamefont {R.~A.}\ \bibnamefont
  {Barton}}, \bibinfo {author} {\bibfnamefont {S.~S.}\ \bibnamefont
  {Verbridge}}, \bibinfo {author} {\bibfnamefont {D.~R.}\ \bibnamefont
  {Southworth}}, \bibinfo {author} {\bibfnamefont {B.}~\bibnamefont {Ilic}},
  \bibinfo {author} {\bibfnamefont {H.~G.}\ \bibnamefont {Craighead}}, \ and\
  \bibinfo {author} {\bibfnamefont {J.~M.}\ \bibnamefont {Parpia}},\ }\href
  {\doibase 10.1103/PhysRevLett.106.047205} {\bibfield  {journal} {\bibinfo
  {journal} {Phys. Rev. Lett.}\ }\textbf {\bibinfo {volume} {106}},\ \bibinfo
  {pages} {047205} (\bibinfo {year} {2011})}\BibitemShut {NoStop}%
\bibitem [{\citenamefont {Cole}\ \emph {et~al.}(2011)\citenamefont {Cole},
  \citenamefont {Wilson-Rae}, \citenamefont {Werbach}, \citenamefont {Vanner},\
  and\ \citenamefont {Aspelmeyer}}]{Cole2011}%
  \BibitemOpen
  \bibfield  {author} {\bibinfo {author} {\bibfnamefont {G.~D.}\ \bibnamefont
  {Cole}}, \bibinfo {author} {\bibfnamefont {I.}~\bibnamefont {Wilson-Rae}},
  \bibinfo {author} {\bibfnamefont {K.}~\bibnamefont {Werbach}}, \bibinfo
  {author} {\bibfnamefont {M.~R.}\ \bibnamefont {Vanner}}, \ and\ \bibinfo
  {author} {\bibfnamefont {M.}~\bibnamefont {Aspelmeyer}},\ }\href
  {http://dx.doi.org/10.1038/ncomms1212} {\bibfield  {journal} {\bibinfo
  {journal} {Nat Commun}\ }\textbf {\bibinfo {volume} {2}},\ \bibinfo {pages}
  {231} (\bibinfo {year} {2011})}\BibitemShut {NoStop}%
\bibitem [{\citenamefont {Lifshitz}\ and\ \citenamefont
  {Roukes}(2000)}]{PhysRevB.61.5600}%
  \BibitemOpen
  \bibfield  {author} {\bibinfo {author} {\bibfnamefont {R.}~\bibnamefont
  {Lifshitz}}\ and\ \bibinfo {author} {\bibfnamefont {M.~L.}\ \bibnamefont
  {Roukes}},\ }\href {\doibase 10.1103/PhysRevB.61.5600} {\bibfield  {journal}
  {\bibinfo  {journal} {Phys. Rev. B}\ }\textbf {\bibinfo {volume} {61}},\
  \bibinfo {pages} {5600} (\bibinfo {year} {2000})}\BibitemShut {NoStop}%
\bibitem [{\citenamefont {Faust}\ \emph {et~al.}(2013)\citenamefont {Faust},
  \citenamefont {Rieger}, \citenamefont {Seitner}, \citenamefont {Kotthaus},\
  and\ \citenamefont {Weig}}]{Faust2013}%
  \BibitemOpen
  \bibfield  {author} {\bibinfo {author} {\bibfnamefont {T.}~\bibnamefont
  {Faust}}, \bibinfo {author} {\bibfnamefont {J.}~\bibnamefont {Rieger}},
  \bibinfo {author} {\bibfnamefont {M.~J.}\ \bibnamefont {Seitner}}, \bibinfo
  {author} {\bibfnamefont {J.~P.}\ \bibnamefont {Kotthaus}}, \ and\ \bibinfo
  {author} {\bibfnamefont {E.~M.}\ \bibnamefont {Weig}},\ }\href
  {http://dx.doi.org/10.1038/nphys2666} {\bibfield  {journal} {\bibinfo
  {journal} {Nat Phys}\ }\textbf {\bibinfo {volume} {9}},\ \bibinfo {pages}
  {485} (\bibinfo {year} {2013})}\BibitemShut {NoStop}%
\bibitem [{\citenamefont {Tielb\"urger}\ \emph {et~al.}(1992)\citenamefont
  {Tielb\"urger}, \citenamefont {Merz}, \citenamefont {Ehrenfels},\ and\
  \citenamefont {Hunklinger}}]{PhysRevB.45.2750}%
  \BibitemOpen
  \bibfield  {author} {\bibinfo {author} {\bibfnamefont {D.}~\bibnamefont
  {Tielb\"urger}}, \bibinfo {author} {\bibfnamefont {R.}~\bibnamefont {Merz}},
  \bibinfo {author} {\bibfnamefont {R.}~\bibnamefont {Ehrenfels}}, \ and\
  \bibinfo {author} {\bibfnamefont {S.}~\bibnamefont {Hunklinger}},\ }\href
  {\doibase 10.1103/PhysRevB.45.2750} {\bibfield  {journal} {\bibinfo
  {journal} {Phys. Rev. B}\ }\textbf {\bibinfo {volume} {45}},\ \bibinfo
  {pages} {2750} (\bibinfo {year} {1992})}\BibitemShut {NoStop}%
\bibitem [{\citenamefont {Pohl}\ \emph {et~al.}(2002)\citenamefont {Pohl},
  \citenamefont {Liu},\ and\ \citenamefont {Thompson}}]{RevModPhys.74.991}%
  \BibitemOpen
  \bibfield  {author} {\bibinfo {author} {\bibfnamefont {R.~O.}\ \bibnamefont
  {Pohl}}, \bibinfo {author} {\bibfnamefont {X.}~\bibnamefont {Liu}}, \ and\
  \bibinfo {author} {\bibfnamefont {E.}~\bibnamefont {Thompson}},\ }\href
  {\doibase 10.1103/RevModPhys.74.991} {\bibfield  {journal} {\bibinfo
  {journal} {Rev. Mod. Phys.}\ }\textbf {\bibinfo {volume} {74}},\ \bibinfo
  {pages} {991} (\bibinfo {year} {2002})}\BibitemShut {NoStop}%
\bibitem [{\citenamefont {Vacher}\ \emph {et~al.}(2005)\citenamefont {Vacher},
  \citenamefont {Courtens},\ and\ \citenamefont {Foret}}]{PhysRevB.72.214205}%
  \BibitemOpen
  \bibfield  {author} {\bibinfo {author} {\bibfnamefont {R.}~\bibnamefont
  {Vacher}}, \bibinfo {author} {\bibfnamefont {E.}~\bibnamefont {Courtens}}, \
  and\ \bibinfo {author} {\bibfnamefont {M.}~\bibnamefont {Foret}},\ }\href
  {\doibase 10.1103/PhysRevB.72.214205} {\bibfield  {journal} {\bibinfo
  {journal} {Phys. Rev. B}\ }\textbf {\bibinfo {volume} {72}},\ \bibinfo
  {pages} {214205} (\bibinfo {year} {2005})}\BibitemShut {NoStop}%
\bibitem [{\citenamefont {Zink}\ \emph {et~al.}(2006)\citenamefont {Zink},
  \citenamefont {Pietri},\ and\ \citenamefont
  {Hellman}}]{PhysRevLett.96.055902}%
  \BibitemOpen
  \bibfield  {author} {\bibinfo {author} {\bibfnamefont {B.~L.}\ \bibnamefont
  {Zink}}, \bibinfo {author} {\bibfnamefont {R.}~\bibnamefont {Pietri}}, \ and\
  \bibinfo {author} {\bibfnamefont {F.}~\bibnamefont {Hellman}},\ }\href
  {\doibase 10.1103/PhysRevLett.96.055902} {\bibfield  {journal} {\bibinfo
  {journal} {Phys. Rev. Lett.}\ }\textbf {\bibinfo {volume} {96}},\ \bibinfo
  {pages} {055902} (\bibinfo {year} {2006})}\BibitemShut {NoStop}%
\bibitem [{\citenamefont {Arcizet}\ \emph {et~al.}(2009)\citenamefont
  {Arcizet}, \citenamefont {Rivi\`ere}, \citenamefont {Schliesser},
  \citenamefont {Anetsberger},\ and\ \citenamefont
  {Kippenberg}}]{PhysRevA.80.021803}%
  \BibitemOpen
  \bibfield  {author} {\bibinfo {author} {\bibfnamefont {O.}~\bibnamefont
  {Arcizet}}, \bibinfo {author} {\bibfnamefont {R.}~\bibnamefont {Rivi\`ere}},
  \bibinfo {author} {\bibfnamefont {A.}~\bibnamefont {Schliesser}}, \bibinfo
  {author} {\bibfnamefont {G.}~\bibnamefont {Anetsberger}}, \ and\ \bibinfo
  {author} {\bibfnamefont {T.~J.}\ \bibnamefont {Kippenberg}},\ }\href
  {\doibase 10.1103/PhysRevA.80.021803} {\bibfield  {journal} {\bibinfo
  {journal} {Phys. Rev. A}\ }\textbf {\bibinfo {volume} {80}},\ \bibinfo
  {pages} {021803} (\bibinfo {year} {2009})}\BibitemShut {NoStop}%
\bibitem [{\citenamefont {Suh}\ \emph {et~al.}(2013)\citenamefont {Suh},
  \citenamefont {Weinstein},\ and\ \citenamefont {Schwab}}]{suh:052604}%
  \BibitemOpen
  \bibfield  {author} {\bibinfo {author} {\bibfnamefont {J.}~\bibnamefont
  {Suh}}, \bibinfo {author} {\bibfnamefont {A.~J.}\ \bibnamefont {Weinstein}},
  \ and\ \bibinfo {author} {\bibfnamefont {K.~C.}\ \bibnamefont {Schwab}},\
  }\href {\doibase 10.1063/1.4816428} {\bibfield  {journal} {\bibinfo
  {journal} {Applied Physics Letters}\ }\textbf {\bibinfo {volume} {103}},\
  \bibinfo {eid} {052604} (\bibinfo {year} {2013})}\BibitemShut {NoStop}%
\bibitem [{\citenamefont {Wilson-Rae}(2013)}]{Wilson-Rae2013}%
  \BibitemOpen
  \bibfield  {author} {\bibinfo {author} {\bibfnamefont {I.}~\bibnamefont
  {Wilson-Rae}},\ }\href@noop {} {\enquote {\bibinfo {title} {Private
  communication},}\ } (\bibinfo {year} {2013})\BibitemShut {NoStop}%
\bibitem [{\citenamefont {{Faust}}\ \emph {et~al.}(2012)\citenamefont
  {{Faust}}, \citenamefont {{Krenn}}, \citenamefont {{Manus}}, \citenamefont
  {{Kotthaus}},\ and\ \citenamefont {Weig}}]{Faust2012}%
  \BibitemOpen
  \bibfield  {author} {\bibinfo {author} {\bibfnamefont {T.}~\bibnamefont
  {{Faust}}}, \bibinfo {author} {\bibfnamefont {P.}~\bibnamefont {{Krenn}}},
  \bibinfo {author} {\bibfnamefont {S.}~\bibnamefont {{Manus}}}, \bibinfo
  {author} {\bibfnamefont {J.~P.}\ \bibnamefont {{Kotthaus}}}, \ and\ \bibinfo
  {author} {\bibfnamefont {E.~M.}\ \bibnamefont {Weig}},\ }\href
  {http://dx.doi.org/10.1038/ncomms1723} {\bibfield  {journal} {\bibinfo
  {journal} {Nature Commun.}\ }\textbf {\bibinfo {volume} {3}},\ \bibinfo
  {pages} {728} (\bibinfo {year} {2012})}\BibitemShut {NoStop}%
\bibitem [{\citenamefont {Unterreithmeier}\ \emph {et~al.}(2009)\citenamefont
  {Unterreithmeier}, \citenamefont {Weig},\ and\ \citenamefont
  {Kotthaus}}]{Unterreithmeier2009}%
  \BibitemOpen
  \bibfield  {author} {\bibinfo {author} {\bibfnamefont {Q.~P.}\ \bibnamefont
  {Unterreithmeier}}, \bibinfo {author} {\bibfnamefont {E.~M.}\ \bibnamefont
  {Weig}}, \ and\ \bibinfo {author} {\bibfnamefont {J.~P.}\ \bibnamefont
  {Kotthaus}},\ }\href {http://dx.doi.org/10.1038/nature07932} {\bibfield
  {journal} {\bibinfo  {journal} {Nature}\ }\textbf {\bibinfo {volume} {458}},\
  \bibinfo {pages} {1001} (\bibinfo {year} {2009})}\BibitemShut {NoStop}%
\bibitem [{\citenamefont {Rieger}\ \emph {et~al.}(2012)\citenamefont {Rieger},
  \citenamefont {Faust}, \citenamefont {Seitner}, \citenamefont {Kotthaus},\
  and\ \citenamefont {Weig}}]{rieger:103110}%
  \BibitemOpen
  \bibfield  {author} {\bibinfo {author} {\bibfnamefont {J.}~\bibnamefont
  {Rieger}}, \bibinfo {author} {\bibfnamefont {T.}~\bibnamefont {Faust}},
  \bibinfo {author} {\bibfnamefont {M.~J.}\ \bibnamefont {Seitner}}, \bibinfo
  {author} {\bibfnamefont {J.~P.}\ \bibnamefont {Kotthaus}}, \ and\ \bibinfo
  {author} {\bibfnamefont {E.~M.}\ \bibnamefont {Weig}},\ }\href {\doibase
  10.1063/1.4751351} {\bibfield  {journal} {\bibinfo  {journal} {Appl. Phys.
  Lett.}\ }\textbf {\bibinfo {volume} {101}},\ \bibinfo {eid} {103110}
  (\bibinfo {year} {2012})}\BibitemShut {NoStop}%
\bibitem [{\citenamefont {Teufel}\ \emph {et~al.}(2008)\citenamefont {Teufel},
  \citenamefont {Regal},\ and\ \citenamefont
  {Lehnert}}]{1367-2630-10-9-095002}%
  \BibitemOpen
  \bibfield  {author} {\bibinfo {author} {\bibfnamefont {J.~D.}\ \bibnamefont
  {Teufel}}, \bibinfo {author} {\bibfnamefont {C.~A.}\ \bibnamefont {Regal}}, \
  and\ \bibinfo {author} {\bibfnamefont {K.~W.}\ \bibnamefont {Lehnert}},\
  }\href {http://stacks.iop.org/1367-2630/10/i=9/a=095002} {\bibfield
  {journal} {\bibinfo  {journal} {New J. Phys.}\ }\textbf {\bibinfo {volume}
  {10}},\ \bibinfo {pages} {095002} (\bibinfo {year} {2008})}\BibitemShut
  {NoStop}%
\bibitem [{\citenamefont {Bordoni}\ \emph {et~al.}(1961)\citenamefont
  {Bordoni}, \citenamefont {Nuovo},\ and\ \citenamefont
  {Verdini}}]{PhysRev.123.1204}%
  \BibitemOpen
  \bibfield  {author} {\bibinfo {author} {\bibfnamefont {P.~G.}\ \bibnamefont
  {Bordoni}}, \bibinfo {author} {\bibfnamefont {M.}~\bibnamefont {Nuovo}}, \
  and\ \bibinfo {author} {\bibfnamefont {L.}~\bibnamefont {Verdini}},\ }\href
  {\doibase 10.1103/PhysRev.123.1204} {\bibfield  {journal} {\bibinfo
  {journal} {Phys. Rev.}\ }\textbf {\bibinfo {volume} {123}},\ \bibinfo {pages}
  {1204} (\bibinfo {year} {1961})}\BibitemShut {NoStop}%
\bibitem [{\citenamefont {Nowick}\ and\ \citenamefont
  {Berry}(1972)}]{Nowick1972}%
  \BibitemOpen
  \bibfield  {author} {\bibinfo {author} {\bibfnamefont {A.}~\bibnamefont
  {Nowick}}\ and\ \bibinfo {author} {\bibfnamefont {B.}~\bibnamefont {Berry}},\
  }\href {\doibase 10.1002/pol.1973.130110713} {\emph {\bibinfo {title}
  {Anelastic Relaxation in Crystalline Solids}}}\ (\bibinfo  {publisher}
  {Academic Press},\ \bibinfo {address} {New York},\ \bibinfo {year}
  {1972})\BibitemShut {NoStop}%
\bibitem [{\citenamefont {Kappesser}\ \emph {et~al.}(1996)\citenamefont
  {Kappesser}, \citenamefont {Wipf}, \citenamefont {Barnes},\ and\
  \citenamefont {Beaudry}}]{Kappesser1996}%
  \BibitemOpen
  \bibfield  {author} {\bibinfo {author} {\bibfnamefont {B.}~\bibnamefont
  {Kappesser}}, \bibinfo {author} {\bibfnamefont {H.}~\bibnamefont {Wipf}},
  \bibinfo {author} {\bibfnamefont {R.~G.}\ \bibnamefont {Barnes}}, \ and\
  \bibinfo {author} {\bibfnamefont {B.~J.}\ \bibnamefont {Beaudry}},\ }\href
  {http://stacks.iop.org/0295-5075/36/i=5/a=385} {\bibfield  {journal}
  {\bibinfo  {journal} {Europhys. Lett.}\ }\textbf {\bibinfo {volume} {36}},\
  \bibinfo {pages} {385} (\bibinfo {year} {1996})}\BibitemShut {NoStop}%
\bibitem [{\citenamefont {Hutchinson}\ \emph {et~al.}(2004)\citenamefont
  {Hutchinson}, \citenamefont {Truitt}, \citenamefont {Schwab}, \citenamefont
  {Sekaric}, \citenamefont {Parpia}, \citenamefont {Craighead},\ and\
  \citenamefont {Butler}}]{hutchinson:972}%
  \BibitemOpen
  \bibfield  {author} {\bibinfo {author} {\bibfnamefont {A.~B.}\ \bibnamefont
  {Hutchinson}}, \bibinfo {author} {\bibfnamefont {P.~A.}\ \bibnamefont
  {Truitt}}, \bibinfo {author} {\bibfnamefont {K.~C.}\ \bibnamefont {Schwab}},
  \bibinfo {author} {\bibfnamefont {L.}~\bibnamefont {Sekaric}}, \bibinfo
  {author} {\bibfnamefont {J.~M.}\ \bibnamefont {Parpia}}, \bibinfo {author}
  {\bibfnamefont {H.~G.}\ \bibnamefont {Craighead}}, \ and\ \bibinfo {author}
  {\bibfnamefont {J.~E.}\ \bibnamefont {Butler}},\ }\href {\doibase
  10.1063/1.1646213} {\bibfield  {journal} {\bibinfo  {journal} {Applied
  Physics Letters}\ }\textbf {\bibinfo {volume} {84}},\ \bibinfo {pages} {972}
  (\bibinfo {year} {2004})}\BibitemShut {NoStop}%
\bibitem [{\citenamefont {Martin}\ \emph {et~al.}(2010)\citenamefont {Martin},
  \citenamefont {Bassiri}, \citenamefont {Nawrodt}, \citenamefont {Fejer},
  \citenamefont {Gretarsson}, \citenamefont {Gustafson}, \citenamefont {Harry},
  \citenamefont {Hough}, \citenamefont {MacLaren}, \citenamefont {Penn},
  \citenamefont {Reid}, \citenamefont {Route}, \citenamefont {Rowan},
  \citenamefont {Schwarz}, \citenamefont {Seidel}, \citenamefont {Scott},\ and\
  \citenamefont {Woodcraft}}]{Martin2010}%
  \BibitemOpen
  \bibfield  {author} {\bibinfo {author} {\bibfnamefont {I.~W.}\ \bibnamefont
  {Martin}}, \bibinfo {author} {\bibfnamefont {R.}~\bibnamefont {Bassiri}},
  \bibinfo {author} {\bibfnamefont {R.}~\bibnamefont {Nawrodt}}, \bibinfo
  {author} {\bibfnamefont {M.~M.}\ \bibnamefont {Fejer}}, \bibinfo {author}
  {\bibfnamefont {A.}~\bibnamefont {Gretarsson}}, \bibinfo {author}
  {\bibfnamefont {E.}~\bibnamefont {Gustafson}}, \bibinfo {author}
  {\bibfnamefont {G.}~\bibnamefont {Harry}}, \bibinfo {author} {\bibfnamefont
  {J.}~\bibnamefont {Hough}}, \bibinfo {author} {\bibfnamefont
  {I.}~\bibnamefont {MacLaren}}, \bibinfo {author} {\bibfnamefont
  {S.}~\bibnamefont {Penn}}, \bibinfo {author} {\bibfnamefont {S.}~\bibnamefont
  {Reid}}, \bibinfo {author} {\bibfnamefont {R.}~\bibnamefont {Route}},
  \bibinfo {author} {\bibfnamefont {S.}~\bibnamefont {Rowan}}, \bibinfo
  {author} {\bibfnamefont {C.}~\bibnamefont {Schwarz}}, \bibinfo {author}
  {\bibfnamefont {P.}~\bibnamefont {Seidel}}, \bibinfo {author} {\bibfnamefont
  {J.}~\bibnamefont {Scott}}, \ and\ \bibinfo {author} {\bibfnamefont {A.~L.}\
  \bibnamefont {Woodcraft}},\ }\href {\doibase 10.1088/0264-9381/27/22/225020}
  {\bibfield  {journal} {\bibinfo  {journal} {Classical and Quantum Gravity}\
  }\textbf {\bibinfo {volume} {27}},\ \bibinfo {pages} {225020} (\bibinfo
  {year} {2010})}\BibitemShut {NoStop}%
\bibitem [{\citenamefont {Chow}\ \emph {et~al.}(1982)\citenamefont {Chow},
  \citenamefont {Lanford}, \citenamefont {Ke-Ming},\ and\ \citenamefont
  {Rosler}}]{chow:5630}%
  \BibitemOpen
  \bibfield  {author} {\bibinfo {author} {\bibfnamefont {R.}~\bibnamefont
  {Chow}}, \bibinfo {author} {\bibfnamefont {W.~A.}\ \bibnamefont {Lanford}},
  \bibinfo {author} {\bibfnamefont {W.}~\bibnamefont {Ke-Ming}}, \ and\
  \bibinfo {author} {\bibfnamefont {R.~S.}\ \bibnamefont {Rosler}},\ }\href
  {\doibase 10.1063/1.331445} {\bibfield  {journal} {\bibinfo  {journal} {J.
  Appl. Phys.}\ }\textbf {\bibinfo {volume} {53}},\ \bibinfo {pages} {5630}
  (\bibinfo {year} {1982})}\BibitemShut {NoStop}%
\bibitem [{\citenamefont {Shimada}\ \emph {et~al.}(1984)\citenamefont
  {Shimada}, \citenamefont {Matsushita}, \citenamefont {Kuratani},
  \citenamefont {Okamoto}, \citenamefont {Koizumi}, \citenamefont {Tsukuma},\
  and\ \citenamefont {Tsukidate}}]{Shimada1984}%
  \BibitemOpen
  \bibfield  {author} {\bibinfo {author} {\bibfnamefont {M.}~\bibnamefont
  {Shimada}}, \bibinfo {author} {\bibfnamefont {K.}~\bibnamefont {Matsushita}},
  \bibinfo {author} {\bibfnamefont {S.}~\bibnamefont {Kuratani}}, \bibinfo
  {author} {\bibfnamefont {T.}~\bibnamefont {Okamoto}}, \bibinfo {author}
  {\bibfnamefont {M.}~\bibnamefont {Koizumi}}, \bibinfo {author} {\bibfnamefont
  {K.}~\bibnamefont {Tsukuma}}, \ and\ \bibinfo {author} {\bibfnamefont
  {T.}~\bibnamefont {Tsukidate}},\ }\href {\doibase
  10.1111/j.1151-2916.1984.tb09612.x} {\bibfield  {journal} {\bibinfo
  {journal} {Journal of the American Ceramic Society}\ }\textbf {\bibinfo
  {volume} {67}},\ \bibinfo {pages} {C} (\bibinfo {year} {1984})}\BibitemShut
  {NoStop}%
\bibitem [{\citenamefont {Holmes}\ \emph {et~al.}(1998)\citenamefont {Holmes},
  \citenamefont {Gildemeister}, \citenamefont {Richards},\ and\ \citenamefont
  {Kotsubo}}]{holmes:2250}%
  \BibitemOpen
  \bibfield  {author} {\bibinfo {author} {\bibfnamefont {W.}~\bibnamefont
  {Holmes}}, \bibinfo {author} {\bibfnamefont {J.~M.}\ \bibnamefont
  {Gildemeister}}, \bibinfo {author} {\bibfnamefont {P.~L.}\ \bibnamefont
  {Richards}}, \ and\ \bibinfo {author} {\bibfnamefont {V.}~\bibnamefont
  {Kotsubo}},\ }\href {\doibase 10.1063/1.121269} {\bibfield  {journal}
  {\bibinfo  {journal} {Applied Physics Letters}\ }\textbf {\bibinfo {volume}
  {72}},\ \bibinfo {pages} {2250} (\bibinfo {year} {1998})}\BibitemShut
  {NoStop}%
\bibitem [{\citenamefont {Rieger}\ \emph {et~al.}(2013)\citenamefont {Rieger},
  \citenamefont {Isacsson}, \citenamefont {Seitner}, \citenamefont {Kotthaus},\
  and\ \citenamefont {Weig}}]{Rieger2013}%
  \BibitemOpen
  \bibfield  {author} {\bibinfo {author} {\bibfnamefont {J.}~\bibnamefont
  {Rieger}}, \bibinfo {author} {\bibfnamefont {A.}~\bibnamefont {Isacsson}},
  \bibinfo {author} {\bibfnamefont {M.~J.}\ \bibnamefont {Seitner}}, \bibinfo
  {author} {\bibfnamefont {J.~P.}\ \bibnamefont {Kotthaus}}, \ and\ \bibinfo
  {author} {\bibfnamefont {E.~M.}\ \bibnamefont {Weig}},\ }\href@noop {}
  {\bibfield  {journal} {\bibinfo  {journal} {in preparation}\ } (\bibinfo
  {year} {2013})}\BibitemShut {NoStop}%
\end{thebibliography}%

\end{document}